\documentclass[twocolumn,showpacs,preprintnumbers,amsmath,amssymb]{revtex4}
\input psfig.sty

\usepackage{graphicx}
\usepackage{dcolumn}
\usepackage{bm}

\begin{document}


\title{Energy Conditions and Segre Classification of Phantom Fields}


\author{Janilo Santos} \email{janilo@dfte.ufrn.br}
\affiliation{Universidade Federal do Rio Grande do Norte, Departamento de
F\'{\i}sica, C.P. 1641, 59072-970 Natal, RN, Brasil}

\author{J. S. Alcaniz} \email{alcaniz@on.br}
\affiliation{Observat\'orio Nacional, Rua Gal. Jos\'e Cristino 77,
20921-400 Rio de Janeiro - RJ, Brasil}

\date{\today}

\begin{abstract}

Recent discoveries in the field of observational cosmology
have provided increasing evidence that the Universe is undergoing a late
time acceleration, which has also stimulated speculations on the nature
of the dark component responsible for such a phenomenon. Among several
candidates discussed in the current literature, phantom fields, an
exotic scalar field with a negative kinetic term and that violates most
of the classical energy conditions, appear as a real possibility
according to recent observational analysis. In this paper we examine the
invariant characterization for  the energy-momentum tensor of phantom
fields through the Segre algebraic classification in the framework of
general relativity.  We also discuss some constraints which are imposed on
the values of $V(\phi)$ from the classical energy conditions.

\end{abstract}

\maketitle


\section{Introduction}

The algebraic classification of symmetric second-order tensors
locally defined on a 4-dimensional Lorentzian manifold,
such as the Ricci tensor $R_{ab}$, the Einstein tensor $G_{ab}$ and
the energy-momentum tensor $T_{ab}$, is known as Segre
classification~\cite{Hall-1}.
The major idea underlying most of classifications in science
is the concept of equivalence. Clearly the objects may be
classified in different ways according to the criteria one chooses
to group them into equivalence classes. There are, however, criteria
that prove to be more important than others. The great
appeal of Segre classification in general relativity is that it
incorporates, ab initio, the Lorentzian character of spacetime.
It is of interest in several contexts
such as, for example, in understanding purely geometrical features of
space-times \cite{Churchill}, in classifying and
interpreting matter fields in general
relativity \cite{Hall-2} and in higher
dimensional theories~\cite{5Djmp}
(e.g., 5-D brane-worlds~\cite{Note}) or still as part of the
procedure for checking whether apparently different space-times
are in fact locally the same up to coordinate transformations
(equivalence problem~\cite{Karlhede}). Because of Einstein's equations
$G_{ab}=\kappa T_{ab}$ the Einsten tensor (and so $R_{ab}$)
has the same algebraic classification as the energy-momentum tensor.
This can be used to decide, on base of generic features such as
the Segre class, which energy-momentum tensors do couple to a
given geometry.

A great deal of difficulty in Segre classifying second order
tensors in the context of general relativity is that
while $G_{ab}$ and $R_{ab}$ are universal functions of the space-time
geometry, $T_{ab}$ depends on the symmetries of the model as well as
on the particular type or combination of matter fields present in the
cosmological scenario.
In this regard, a powerful way of imposing physical constraints and
searching realistic forms for the energy-momentum tensor is through
cosmological observations. The high degree of isotropy observed in the
Cosmic Microwave Background Radiation \cite{wmap}, for instance,
restricts the general form of $T_{ab}$ while results from distance
measurements using type Ia supernovae (which constitute the most direct
evidence for the current cosmic acceleration) \cite{riess} impose
important constraints on the physical quantities of the cosmic fluid.

In light of an impressive convergence of observational results, several
groups have recently tested the viability of different matter fields
(or, equivalently, different forms of $T_{ab}$) as a realistic
description for the matter content of our Universe.
The two favorite candidates are the energy density stored on the true
vacuum state of all existing fields in the Universe, i.e., the cosmological
constant ($T^{\Lambda}_{ab} = \Lambda g_{ab}$) and the potential energy
density $V(\phi)$ associated with a dynamical scalar field $\phi$,
usually called dynamical dark energy (see, e.g., \cite{review} for a
recent review on this topic). Physically, the role of such a dynamical
dark energy can be played by any scalar field violating
the so-called strong energy condition (see Sec. III for a discussion).
Two possibilities, however, have been exhaustively explored in the
current literature, namely, quintessence scalar fields with positive
kinetic term and an equation-of-state parameter $w \geq -1$ \cite{quint},
and phantom fields with
negative kinetic term and $w < -1$ \cite{phantom,phantom1,Singh}.
In terms of the parameter $w$, the former case can
be seen as a very general scenario, which includes cold dark matter
models with a cosmological constant ($\Lambda$CDM), $w = -1$, and
cosmologies dominated by topological defects (e.g., domain walls,
strings and textures) for which $w = -\frac{n}{3}$, $n$ being the
dimension of the defect. The latter case in turn is characterized by
strange properties such as, for instance, the fact that its energy
density increases with the expansion of the Universe (in contrast with
quintessence fields); the possibility of a rip-off of
large and small scale structures of matter; a possible occurrence of future
curvature singularity, etc. \cite{phantom1}. Although having these unusual
characteristics, a phantom type behavior is predicted by several scenarios,
e.g., kinetically driven models \cite{kine} and some versions of brane world
cosmologies \cite{brane}. From the observational point of view, phantom
dark energy is found to be compatible with most of the classical
cosmological tests and provide a better fit to type Ia supernovae
observations than do $\Lambda$CDM or quintessence scenarios ($w > -1$).
Therefore this means that, although exotic, phantom fields may be the
dominant form of energy in our universe.

In this paper we examine the invariant characterization for  the
energy-momentum tensor of phantom fields through the Segre algebraic
classification in the framework of general relativity. In Sec. II, by using
real null tetrad technique, we show that phantom fields can be
classified in two different subclasses of equivalence which are represented
by Segre types $[1,(111)]$ and $[(1,111)]$. In Sec. III we present
the so-called energy conditions of general relativity for the
whole Segre class $[1,111]$ and also examine the further
restrictions they impose on the values of the phantom field potential
$V(\phi)$. We end this paper by summarizing the main results in the
conclusion Section.


\section{Segre Classification}

A classification of a generic symmetric second order tensor
$T_{ab}$ can be cast in terms of the eigenvalue problem
\begin{equation}
  (\, T^{a}_{\ b} \,- \, \lambda \,\, \delta^{a}_{\ b} \,)
\; v^{b} \, = 0                             \label{eigenprob}
\end{equation}
where $\lambda$ are eigenvalues, $v^{a}$ are eigenvectors, and
the mixed tensor $T^{a}_{\ b}$ may be thought of as a linear
map $T_{p}(M) \mapsto T_{p}(M)$. $M$ is a real 4-dimensional space-time
manifold locally endowed with a Lorentzian metric of signature
$(- + + + )$, $T_{p}(M)$ denotes the tangent space to $M$ at a point
$p \in M$ and latin indices range from 0 to 3. Because of the
Lorentzian character of the metric the mixed form of the energy-momentum
tensor is no more symmetric and $T^a_b$ may not have a diagonal matrix
representation, i.e., it need not have four linearly independent
eigenvectors. However, using the {\em Jordan canonical
forms} of the matrix of an operator, and imposing
the Lorentzian character of the metric on $M$, it has been
shown~\cite{Hall-1,Hall-2} (for detailed calculations in $n\geq 5$
dimensional space-times and a review on this topic see~\cite{BJP})
that any energy-momentum tensor defined on $T_{p}(M)$ reduces to one
of the four canonical forms:
\begin{equation}  \label{segre-classes}
\begin{array}{cll}  
{[}1,111]  & T_{ab} = & 2\,\sigma_1\,l_{(a}m_{b)}
+ \sigma_2\,(l_{a}l_{b} + m_{a}m_{b}) \\
&  & + \sigma_3\,x_{a}x_{b}  +  \sigma_4\,y_{a}y_{b}   \\
\\
{[}211]  & T_{ab} = & 2\,\sigma_1\,l_{(a}m_{b)} \pm l_{a}l_{b} +
\sigma_2\,x_{a}x_{b} \\
&  & + \sigma_3\,y_{a}y_{b}   \\
\\
{[}31]   & T_{ab} = & 2\,\sigma_1\,l_{(a}m_{b)} + 2\,l_{(a}x_{b)} +
\sigma_1\,x_{a}x_{b} \\
&  & + \sigma_2\,y_{a}y_{b}  \\
\\
{[}z\,\bar{z}\,11]& T_{ab} = & 2\,\sigma_1\,l_{(a}m_{b)} +
\sigma_2\,(l_{a}l_{b} - m_{a}m_{b}) \\
&  & + \sigma_3\,x_{a}x_{b} + \sigma_4\,y_{a}y_{b}\;,
\end{array}
\end{equation}
where $\sigma_1, \cdots ,\sigma_4 \in \mathbb{R}$, having
different values for different formulae,
and in the Segre class $[z\,\bar{z}\,11]$ $\sigma_2\, \neq 0$. In
the above "catalog" the first column is the notation commonly used
for indicating the Segre class, which is a list $[r_1r_2\cdots r_n]$
of the dimensions of the Jordan blocks of the corresponding Jordan
canonical matrix. The basis vectors $\{l,m,x,y\}$ form a real null
tetrad basis such that the only nonvanishing inner products are
\begin{equation}
l^{a}m_{a} = x^{a}x_{a} = y^{a}y_{a} = 1\,. \label{inerp}
\end{equation}
This basis is constructed from the preferred directions intrinsically
defined by the tensor, i.e., from the Jordan basis.
The Segre class $[1,111]$ distinguishes tensors that
have diagonal matrix representation and is the unique that admit a
timelike eigenvector. The comma in this case is used to separate
timelike from spacelike eigenvectors.  Regardless of the dimension of a
Jordan block, there is only one eigenvector associated to each block,
and the eigenvector associated to a block of dimension $r>1$ is a null
vector~\cite{5Djmp}. Energy-momentum tensors which belong to Segre
class $[211]$, for example, have only three linearly independent
eigenvectors (one of which is a null vector), and its characteristic
polynomial obtained from (\ref{eigenprob}) has three roots:
one of multiplicity 2 and the others of multiplicity 1.
So, the digits inside the brackets give also the
multiplicity of the real eigenvalues, the complex one being taken
into account by Segre class $[z\,\bar{z}\,11]$, where the $z\,\bar{z}$
refers to a pair of complex eigenvalues. Degeneracy amongst eigenvalues in
different Jordan blocks will be indicated by enclosing the corresponding
digits inside round brackets, as in $[1,(111)]$, which indicate that
three out of the four eigenvalues are degenerate. So, to each
Segre class there may be several subclasses depending on the
degeneracies of the eigenvalues.


\subsection{Segre Classification of Phantom Fields}

The energy-momentum tensor for the phantom field has the
form~\cite{Singh}
\begin{equation} \label{phantom-stress}
T_{ab}=-\phi_a\phi_b + g_{ab}\left[
\frac{1}{2}\,g^{cd}\phi_c\phi_d - V(\phi) \right]\,,
\end{equation}
where $V(\phi)$ is the phantom potential, $\phi_{a} \equiv\phi_{;a}$
and the semicolon denotes
covariant derivative. Here we assume that $\phi=\phi(t)$ is a
function of time alone evolving in an isotropic and homogeneous
space-time, so $g_{ab}\phi^a\phi^b =-\dot{\phi}^2$, that is, $\phi^a$
is a time-like vector (indeed an eigenvector of $T_{ab}$).
In this case it is always possible to find
out two null vectors $l_{a}$ and $m_{a}$ such that $l^{a}m_{a} = 1$,
and  $\phi_{a}$ can be written as
\begin{equation} \label{timelike}
\phi_a=\frac{\dot{\phi}}{\sqrt{2}}(l_a - m_a)\,.
\end{equation}
Besides, we choose two spacelike vectors $x_{a}$ and $y_{a}$,
belonging to the 2-space orthogonal to the 2-space generated
by $l_{a}$ and $m_{a}$, so as to form the real null tetrad
basis $\{l,m,x,y\}$ defined by (\ref{inerp}). In terms of this
basis the metric tensor is written as
\begin{equation} \label{metric}
g_{ab}=2l_{(a}m_{b)} + x_ax_b + y_ay_b
\end{equation}
where the round brackets indicates symmetrization.
Taking into account equations
(\ref{timelike}) and (\ref{metric}), the canonical form for the
energy-momentum tensor (\ref{phantom-stress}) is then written as
\begin{eqnarray}  \label{phantom-segre}
T_{ab} & =  & 2\,\sigma_{1} \, l_{(a}m_{b)} + \sigma_{2} \, (l_{a}l_{b}
+ m_{a}m_{b}) \nonumber \\ & &
 + \sigma_{3} \, (x_{a}x_{b} + y_{a}y_{b}) \,,
\end{eqnarray}
where
\begin{eqnarray} \label{phantom-sigma}
\sigma_{1} = -V(\phi)\,, \;\;\;\;
\sigma_{2} = -\frac{1}{2}\dot{\phi}^2\,, \;\;\;\;
\sigma_{3} = -\frac{1}{2}\dot{\phi}^{2} - V(\phi)\,,
\end{eqnarray}
indicating that phantom fields belong to Segre class $[1,111]$
(see Eq. (\ref{segre-classes})) for $\sigma_3=\sigma_4$. In order
to find the subclasses we now determine its eigenvalues and
corresponding eigenvectors. For the sake of brevity we present our
results without going into details of calculations, which can be
easily verified from equation (\ref{phantom-segre}) and the
expressions below. We find that the set of linearly independent
eigenvectors and associated eigenvalues of $T^{a}_{\ b}$ is given by
\begin{eqnarray}
\begin{array}{rl}
l^{a} - m^{a} \; \longrightarrow & \;\;\:[\frac{1}{2}\dot{\phi}^2-V(\phi)]\\
l^{a} + m^{a} \; \longrightarrow & -[ \frac{1}{2}\dot{\phi}^2
+ V(\phi)]  \\
        x^{a} \; \longrightarrow & - [ \frac{1}{2}\dot{\phi}^2
+ V(\phi)] \\
        y^{a} \; \longrightarrow & -[ \frac{1}{2}\dot{\phi}^2
+ V(\phi)] \,.
\end{array}
\end{eqnarray}
Note that three out of the four eigenvalues are degenerate, making it
apparent that the corresponding Segre subclass is
$[1,(111)]$. This is the same Segre type as that for a perfect
fluid~\cite{KSMH} (we shall comment this latter).
We emphasize that this classification is independent of the
functional form of the potential $V(\phi)$ as well as the time
derivative $\dot{\phi}$ except for $\phi=\,$cte. In this latter case the
Segre subclass for the phantom field is $[(1,111)]$ with $-V(\phi)$ a
four-time-degenerate eigenvalue. This is the same Segre subclass
of energy-momentum tensors of $\Lambda-$term type (cosmological
constant)~\cite{KSMH}. Since a $\Lambda-$term can always be
incorporated into an energy-momentum tensor of the perfect fluid,
we shall banish this subclass from further consideration.


\section{Classical Energy Conditions for Phantom Fields}

In this Section we investigate possible constraints that the classical
energy conditions may impose on the values of the potential $V(\phi)$
as well as on $\dot{\phi}$ for phantom fields. Restrictions imposed by
energy conditions on energy-momentum tensors of general relativity
theory for matter fields like those represented by Segre classes
$[211]$, $[31]$ and $[z\bar{z}11]$ in Eq. (\ref{segre-classes}),
have been presented in the literature~\cite{Hall-1,Hawking-Ellis}.
As we have found the Segre specializations
$[1,(111)]$ and $[(1,111)]$ for phantom fields, we focus our
attention here only on the Segre class $[1,111]$ of Eq.
(\ref{segre-classes}). The most common energy conditions are
\cite{Hawking-Ellis,Visser} (see also \cite{Visser-Barcelo}
for a recent discussion on this topic):
\begin{itemize}
\item[(i)] The null energy condition (NEC). NEC states that
$T_{ab}n^an^b\geq 0$ for null vectors $n^a\in T_{p}(M)$ which, for
Segre class $[1,111]$, is equivalent to require that
$\sigma_2 - \sigma_1 + \sigma_{\alpha}\geq 0$ ($\alpha=3,\,4$).

\item[(ii)] The weak energy condition (WEC). WEC states that
$T_{ab}t^at^b\geq 0$ for timelike vectors $t^a\in T_{p}(M)$. This
will also imply, by continuity, the NEC. The WEC for Segre class
$[1,111]$ means that $\sigma_2 - \sigma_1\geq 0$.

\item[(iii)] The strong energy condition (SEC). SEC is the assertion
that for any timelike vector $(T_{ab} - T/2g_{ab})t^at^b\geq 0$,
where $T$ is the trace of $T_{ab}$. If $T_{ab}$ belongs to the Segre
class $[1,111]$ then we must have
$2\sigma_2+\sigma_3+\sigma_4\geq 0$.

\item[(iv)] The dominant energy condition (DEC). DEC requires that
$T_{ab}t^at^b\geq 0$ for timelike vectors $t^a\in T_{p}(M)$ and the
additional requirement that $T_{ab}t^b$ be a non-spacelike vector.
By continuity this will also hold for null vectors
$n^a\in T_{p}(M)$.
For energy-momentum tensors of Segre class $[1,111]$ this requires
that $\sigma_2-\sigma_1\geq 0$, and
$\sigma_1-\sigma_2\leq \sigma_{\alpha}\leq \sigma_2-\sigma_1$
($\alpha=3,\,4$).
\end{itemize}
\begin{figure}
\centerline{\psfig{figure=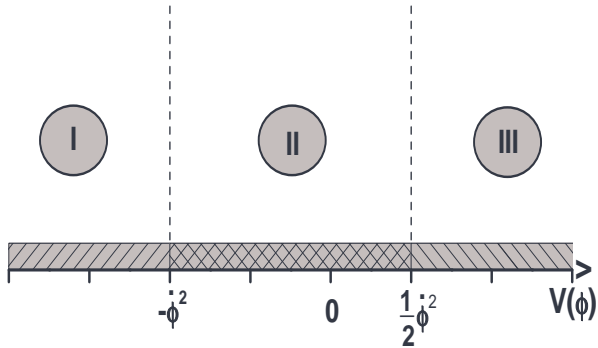,width=3.5truein,height=2.0truein,angle=0}}
\caption{Energy conditions constraints on phantom fields potential.
 In interval {\bf{(I)}} phantom fields do not violate SEC but do
violate WEC. In interval {\bf{(II)}} both (SEC and WEC) are violated
while in region {\bf{(III)}} phantom fields do not violate WEC but
do violate SEC.}
\end{figure}
When imposed on the energy-momentum tensor of the phantom field these energy
conditions require:
\[
\begin{array}{lrr}
\mbox{\bf NEC} & \Rightarrow & -\frac{1}{2}\dot{\phi}^2\geq 0  \\
\mbox{\bf WEC} & \Rightarrow & -\frac{1}{2}\dot{\phi}^2+V(\phi)\geq 0 \\
\mbox{\bf SEC} & \Rightarrow & \dot{\phi}^2 +V(\phi)\leq 0 \\
\mbox{\bf DEC} & \Rightarrow & \quad \quad \quad \quad -\dot{\phi}^2\geq 0
\quad \mbox{and} \quad V(\phi)\geq 0
\end{array}
\]
From the above conditions, it should be noticed that although the Segre
subclass $[1,(111)]$ for the
phantom field is consistent with the Lorentzian signature of the
metric tensor $g_{ab}$, imposition of NEC, as well as DEC, generates at
least one contradiction in the sense that the timelike character
($-\dot{\phi}^2=g_{ab}\phi^a\phi^b$) of the vector $\phi^a$ is violated.
Maintaining the Lorentzian timelike character of $\phi^a$ WEC
can be preserved only for positive potentials since that
$V(\phi)\geq \dot{\phi}^2/2$. SEC, on the other hand, cannot
be satisfied unless $V(\phi)$ is negative and $V(\phi) \leq - \dot{\phi}^2$.
There is however an interval ($-\dot{\phi}^2<V(\phi)<
\dot{\phi}^2/2$) in the values of potential for which
neither SEC nor WEC are preserved. Figure 1 shows such
intervals. In interval I phantom fields do not violate SEC but do
violate WEC. In region II both SEC and WEC are violate while in
region III phantom fields do not violate WEC but do violate
SEC. It is worth mentioning that
evolving quintessence and phantom fields with linear-negative potentials
have been recently proposed and studied in light of recent cosmological
observations. As shown in Ref. \cite{Peri}, such models provide a good
fit for the current supernovae data being, therefore, a potential
candidate for dark energy.

At this point, it is important to observe that in our approach the energy
conditions refer to quantities like potentials and/or time
derivatives of the fields as well as invariantly defined vectors.
It is also possible to present the same results in terms of pressures and
density associated to the phantom field. As we have noted,
the Segre subclass $[1,(111)]$ obtained for
the phantom field is the same as that for a perfect fluid.
This means that the energy-momentum tensor for an ideal fluid
can also be written in the form (\ref{phantom-segre}) by identifying
$\sigma_1=(p-\rho)/2$, $\sigma_2=(p+\rho/2)$ and $\sigma_3=p$,
where $p$ and $\rho$ are the pressure and the density of the fluid.


\section{Conclusion}

Motivated by a number of observational results, the idea of phantom fields
as candidate for the dark energy has been largely explored in the current
literature. In this paper we have examined the algebraic classification of
the energy-momentum tensor $T_{ab}$ of such fields for generic potential
terms. We have shown that the phantom $T_{ab}$ belongs to Segre
subclasses $[1,(111)]$ or $[(1,111)]$. As is well known, all
energy-momentum tensors belonging to these Segre subclasses (e.g.,
perfect fluid and $\Lambda-$term type energy-momentum tensors)
couple to Friedmann-Robertson-Walker geometries.
We also have found some constraints which are imposed on the
values of $V(\phi)$ from the classical energy conditions.
Although the SEC is being violated right now, according recent
observational data regarding the accelerating universe, we see
from our analysis that an evolving potential function $V(\phi)$ can,
in principle, be constructed such that this condition is not violated
in the past (interval I of FIG. 1), while in that epoch WEC must be
violated. Although negative potential for phantom fields have
already been proposed, the transitions from interval I to interval II
and to III demand some studies. Our approach makes also clear that
imposition of NEC, as well as DEC, on phantom fields generates a strong
contradiction about the timelike character of the vector $\phi^a$
composing the energy-momentum tensor.

As discussed earlier, the Segre classification technique can also be applied to the
geometric part of Einstein's equations. In this case, the results could be used to
impose the energy conditions directly, via Segre classes, on the coefficients of a 
given metric such as, for example, the FRW metric. From such an analysis, the production 
of future singularities in an expanding universe without violating the classical energy
conditions could be studied (see, e.g.~\cite{Barrow}). This possibility will be explored 
in a forthcoming communication.


\begin{acknowledgments}

The authors are very grateful to M. J. Rebou\c{c}as for valuable
discussions. JSA thanks the Department of Physics of the
Universidade Federal do Rio Grande do Norte for their hospitality
during the course of this work, and also the financial support by
CNPq (62.0053/01-1-PADCT III/Milenio). JS thanks the support of
PRONEX/CNPq/FAPERN.

\end{acknowledgments}



\end{document}